\documentclass[aps,prl,superscriptaddress,showpacs,floatfix,twocolumn]{revtex4}

\usepackage{graphicx}	
\usepackage{dcolumn}	
\usepackage{multirow}	

\newcommand{\mean}[1]{\left\langle #1 \right\rangle}

\begin{document}

\title{Suppression pattern of neutral pions at high transverse momentum
in Au+Au collisions at $\sqrt{s_{NN}}$=200\,GeV and constraints on medium
transport coefficients}

\newcommand{\abilene}{Abilene Christian University, Abilene, TX 79699, USA}
\newcommand{\banaras}{Department of Physics, Banaras Hindu University, Varanasi 221005, India}
\newcommand{\bnl}{Brookhaven National Laboratory, Upton, NY 11973-5000, USA}
\newcommand{\caucr}{University of California - Riverside, Riverside, CA 92521, USA}
\newcommand{\charlesczech}{Charles University, Ovocn\'{y} trh 5, Praha 1, 116 36, Prague, Czech Republic}
\newcommand{\ciae}{China Institute of Atomic Energy (CIAE), Beijing, People's Republic of China}
\newcommand{\cns}{Center for Nuclear Study, Graduate School of Science, University of Tokyo, 7-3-1 Hongo, Bunkyo, Tokyo 113-0033, Japan}
\newcommand{\colorado}{University of Colorado, Boulder, CO 80309, USA}
\newcommand{\columbia}{Columbia University, New York, NY 10027 and Nevis Laboratories, Irvington, NY 10533, USA}
\newcommand{\czechtech}{Czech Technical University, Zikova 4, 166 36 Prague 6, Czech Republic}
\newcommand{\dapnia}{Dapnia, CEA Saclay, F-91191, Gif-sur-Yvette, France}
\newcommand{\debrecen}{Debrecen University, H-4010 Debrecen, Egyetem t{\'e}r 1, Hungary}
\newcommand{\elte}{ELTE, E{\"o}tv{\"o}s Lor{\'a}nd University, H - 1117 Budapest, P{\'a}zm{\'a}ny P. s. 1/A, Hungary}
\newcommand{\fit}{Florida Institute of Technology, Melbourne, FL 32901, USA}
\newcommand{\fsu}{Florida State University, Tallahassee, FL 32306, USA}
\newcommand{\gsu}{Georgia State University, Atlanta, GA 30303, USA}
\newcommand{\hiroshima}{Hiroshima University, Kagamiyama, Higashi-Hiroshima 739-8526, Japan}
\newcommand{\ihepprot}{IHEP Protvino, State Research Center of Russian Federation, Institute for High Energy Physics, Protvino, 142281, Russia}
\newcommand{\illuiuc}{University of Illinois at Urbana-Champaign, Urbana, IL 61801, USA}
\newcommand{\instpasczech}{Institute of Physics, Academy of Sciences of the Czech Republic, Na Slovance 2, 182 21 Prague 8, Czech Republic}
\newcommand{\isu}{Iowa State University, Ames, IA 50011, USA}
\newcommand{\jinrdubna}{Joint Institute for Nuclear Research, 141980 Dubna, Moscow Region, Russia}
\newcommand{\kaeri}{KAERI, Cyclotron Application Laboratory, Seoul, Korea}
\newcommand{\kek}{KEK, High Energy Accelerator Research Organization, Tsukuba, Ibaraki 305-0801, Japan}
\newcommand{\kfki}{KFKI Research Institute for Particle and Nuclear Physics of the Hungarian Academy of Sciences (MTA KFKI RMKI), H-1525 Budapest 114, POBox 49, Budapest, Hungary}
\newcommand{\korea}{Korea University, Seoul, 136-701, Korea}
\newcommand{\kurchatov}{Russian Research Center ``Kurchatov Institute", Moscow, Russia}
\newcommand{\kyoto}{Kyoto University, Kyoto 606-8502, Japan}
\newcommand{\labllr}{Laboratoire Leprince-Ringuet, Ecole Polytechnique, CNRS-IN2P3, Route de Saclay, F-91128, Palaiseau, France}
\newcommand{\lawllnl}{Lawrence Livermore National Laboratory, Livermore, CA 94550, USA}
\newcommand{\losalamos}{Los Alamos National Laboratory, Los Alamos, NM 87545, USA}
\newcommand{\lpc}{LPC, Universit{\'e} Blaise Pascal, CNRS-IN2P3, Clermont-Fd, 63177 Aubiere Cedex, France}
\newcommand{\lund}{Department of Physics, Lund University, Box 118, SE-221 00 Lund, Sweden}
\newcommand{\muenster}{Institut f\"ur Kernphysik, University of Muenster, D-48149 Muenster, Germany}
\newcommand{\myongji}{Myongji University, Yongin, Kyonggido 449-728, Korea}
\newcommand{\nagasaki}{Nagasaki Institute of Applied Science, Nagasaki-shi, Nagasaki 851-0193, Japan}
\newcommand{\newmex}{University of New Mexico, Albuquerque, NM 87131, USA }
\newcommand{\nmsu}{New Mexico State University, Las Cruces, NM 88003, USA}
\newcommand{\ornl}{Oak Ridge National Laboratory, Oak Ridge, TN 37831, USA}
\newcommand{\orsay}{IPN-Orsay, Universite Paris Sud, CNRS-IN2P3, BP1, F-91406, Orsay, France}
\newcommand{\peking}{Peking University, Beijing, People's Republic of China}
\newcommand{\pnpi}{PNPI, Petersburg Nuclear Physics Institute, Gatchina, Leningrad region, 188300, Russia}
\newcommand{\riken}{RIKEN, The Institute of Physical and Chemical Research, Wako, Saitama 351-0198, Japan}
\newcommand{\rikjrbrc}{RIKEN BNL Research Center, Brookhaven National Laboratory, Upton, NY 11973-5000, USA}
\newcommand{\rikkyo}{Physics Department, Rikkyo University, 3-34-1 Nishi-Ikebukuro, Toshima, Tokyo 171-8501, Japan}
\newcommand{\saispbstu}{Saint Petersburg State Polytechnic University, St. Petersburg, Russia}
\newcommand{\saopaulo}{Universidade de S{\~a}o Paulo, Instituto de F\'{\i}sica, Caixa Postal 66318, S{\~a}o Paulo CEP05315-970, Brazil}
\newcommand{\seoulnat}{System Electronics Laboratory, Seoul National University, Seoul, Korea}
\newcommand{\stonybrkc}{Chemistry Department, Stony Brook University, Stony Brook, SUNY, NY 11794-3400, USA}
\newcommand{\stonycrkp}{Department of Physics and Astronomy, Stony Brook University, SUNY, Stony Brook, NY 11794, USA}
\newcommand{\subatech}{SUBATECH (Ecole des Mines de Nantes, CNRS-IN2P3, Universit{\'e} de Nantes) BP 20722 - 44307, Nantes, France}
\newcommand{\tenn}{University of Tennessee, Knoxville, TN 37996, USA}
\newcommand{\titech}{Department of Physics, Tokyo Institute of Technology, Oh-okayama, Meguro, Tokyo 152-8551, Japan}
\newcommand{\tsukuba}{Institute of Physics, University of Tsukuba, Tsukuba, Ibaraki 305, Japan}
\newcommand{\vandy}{Vanderbilt University, Nashville, TN 37235, USA}
\newcommand{\waseda}{Waseda University, Advanced Research Institute for Science and Engineering, 17 Kikui-cho, Shinjuku-ku, Tokyo 162-0044, Japan}
\newcommand{\weizmann}{Weizmann Institute, Rehovot 76100, Israel}
\newcommand{\yonsei}{Yonsei University, IPAP, Seoul 120-749, Korea}
\affiliation{\abilene}
\affiliation{\banaras}
\affiliation{\bnl}
\affiliation{\caucr}
\affiliation{\charlesczech}
\affiliation{\ciae}
\affiliation{\cns}
\affiliation{\colorado}
\affiliation{\columbia}
\affiliation{\czechtech}
\affiliation{\dapnia}
\affiliation{\debrecen}
\affiliation{\elte}
\affiliation{\fit}
\affiliation{\fsu}
\affiliation{\gsu}
\affiliation{\hiroshima}
\affiliation{\ihepprot}
\affiliation{\illuiuc}
\affiliation{\instpasczech}
\affiliation{\isu}
\affiliation{\jinrdubna}
\affiliation{\kaeri}
\affiliation{\kek}
\affiliation{\kfki}
\affiliation{\korea}
\affiliation{\kurchatov}
\affiliation{\kyoto}
\affiliation{\labllr}
\affiliation{\lawllnl}
\affiliation{\losalamos}
\affiliation{\lpc}
\affiliation{\lund}
\affiliation{\muenster}
\affiliation{\myongji}
\affiliation{\nagasaki}
\affiliation{\newmex}
\affiliation{\nmsu}
\affiliation{\ornl}
\affiliation{\orsay}
\affiliation{\peking}
\affiliation{\pnpi}
\affiliation{\riken}
\affiliation{\rikjrbrc}
\affiliation{\rikkyo}
\affiliation{\saispbstu}
\affiliation{\saopaulo}
\affiliation{\seoulnat}
\affiliation{\stonybrkc}
\affiliation{\stonycrkp}
\affiliation{\subatech}
\affiliation{\tenn}
\affiliation{\titech}
\affiliation{\tsukuba}
\affiliation{\vandy}
\affiliation{\waseda}
\affiliation{\weizmann}
\affiliation{\yonsei}
\author{A.~Adare}	\affiliation{\colorado}
\author{S.~Afanasiev}	\affiliation{\jinrdubna}
\author{C.~Aidala}	\affiliation{\columbia}
\author{N.N.~Ajitanand}	\affiliation{\stonybrkc}
\author{Y.~Akiba}	\affiliation{\riken} \affiliation{\rikjrbrc}
\author{H.~Al-Bataineh}	\affiliation{\nmsu}
\author{J.~Alexander}	\affiliation{\stonybrkc}
\author{A.~Al-Jamel}	\affiliation{\nmsu}
\author{K.~Aoki}	\affiliation{\kyoto} \affiliation{\riken}
\author{L.~Aphecetche}	\affiliation{\subatech}
% ...
\collaboration{PHENIX Collaboration} \noaffiliation

\date{\today}

\begin{abstract}

For Au+Au collisions at 200 GeV we measure neutral pion production with 
good statistics for transverse momentum, $p_{\rm T}$, up to 20\,GeV/$c$.  
A fivefold suppression is found, which is essentially constant for 
5$<$$p_{\rm T}$$<$20\,GeV/$c$. Experimental uncertainties are small enough 
to constrain any model-dependent parameterization for the transport 
coefficient of the medium, e.g. $\mean{\hat{q}}$ in the parton quenching 
model.  The spectral shape is similar for all collision classes, and the 
suppression does not saturate in Au+Au collisions; instead, it increases 
proportional to the number of participating nucleons, as ${N_{\rm 
part}}^{2/3}$.

\end{abstract}

	


\maketitle


%
%

Large transverse momentum ($p_{\rm T}$) hadrons originate primarily from
the fragmentation of hard scattered quarks or gluons.
In high energy p+p collisions this is well
described in the framework of perturbative QCD~\cite{deFlorian:2005yj}.
In ultra-relativistic heavy ion collisions such 
hard scatterings occur in the early phase of the reaction, and 
the transiting partons serve as probes of the strongly interacting
medium produced in the collisions. Lattice QCD predicts a phase
transition to 
a plasma of deconfined quarks and gluons,
which induces gluon radiation
from the scattered parton and depletes hadron production at high $p_{\rm T}$ 
(``jet quenching'')~\cite{Gyulassy:1990ye,Wang:1991xy}.
The measurements in Au+Au collisions at RHIC showed suppressed hadron
yields in central collisions \cite{Adcox:2001jp} as
predicted~\cite{Baier:1996sk,Gyulassy:2000fs},
and motivated an advanced theoretical study of radiative energy loss
using different approaches~\cite{Arnold:2002ja}.

All the energy loss models must incorporate space-time evolution of the
medium, as it is not static, as well as the initial distribution of the
partons throughout the medium. Models generally also include an input
parameter for the medium density and/or the coupling. Different 
assumptions in the various models lead to similar descriptions of the 
$\pi^0$ suppression with different model-dependent 
parameters~\cite{Renk:2006pr,Baier:2007jh}. 
For instance, the Parton Quenching model (PQM) is a Monte Carlo using the
quenching weights from BDMPS~\cite{Baier:1996sk} that combines the coupling
strength with the color-charge density to create a single transport
coefficient, often referred to as
$\mean{\hat{q}}$~\cite{Dainese:2004te,Loizides:2006cs}, which gives the
average squared transverse momentum transferred from the medium to the
parton per mean free path. 

Measurement of identified particles up to the highest possible $p_{\rm T}$,
establishing the magnitude, $p_{\rm T}$ and centrality dependence of the
suppression pattern, is crucial to constrain the theoretical
models and separate contributions of initial and final state effects
from the energy loss mechanism. Collision centrality is related to
the average pathlength of the parton in the medium. 
The suppression of $\pi^0$ puts important constraints on calculations of
the energy loss, as neutral pions can be identified up to very high $p_{\rm T}$.
Whereas it has also been shown that the di-hadron suppression at high $p_{\rm T}$
may be somewhat more sensitive than single hadron suppression 
to the medium opacity~\cite{Zhang:2007ja}, 
such improvement is contingent upon the theoretical and
experimental, statistical and systematic uncertainties.


This Letter reports on the measurement of neutral pions up to
$p_{\rm T}$=20\,GeV/$c$ in Au+Au collisions at $\sqrt{s_{NN}}$=200\,GeV at RHIC, 
using the high statistics collected in Run-4.
Based upon the data 
we extract the $\mean{\hat{q}}$ parameter of the PQM model for the
most central collisions.




%
%
The analysis used $1.03 \times 10^9$ minimum bias events
taken by the PHENIX experiment~\cite{Adcox:2003zm}.
%
%
Collision centrality was determined from the correlation between the
number of charged particles detected in the Beam-Beam Counters
(BBC, $3.0$$<$$|\eta|$$<$$3.9$) and the energy measured in the Zero Degree
Calorimeters (ZDC). A Glauber model Monte Carlo
with a simulation of the BBC and ZDC responses was used to estimate the
associated average number of participating nucleons ($\mean{N_{\rm part}}$) 
and binary nucleon-nucleon collisions ($\mean{N_{\rm coll}}$) for each 
centrality bin~\cite{Miller:2007ri}.


\begin{figure*}[t]
\includegraphics[width=0.8\linewidth]{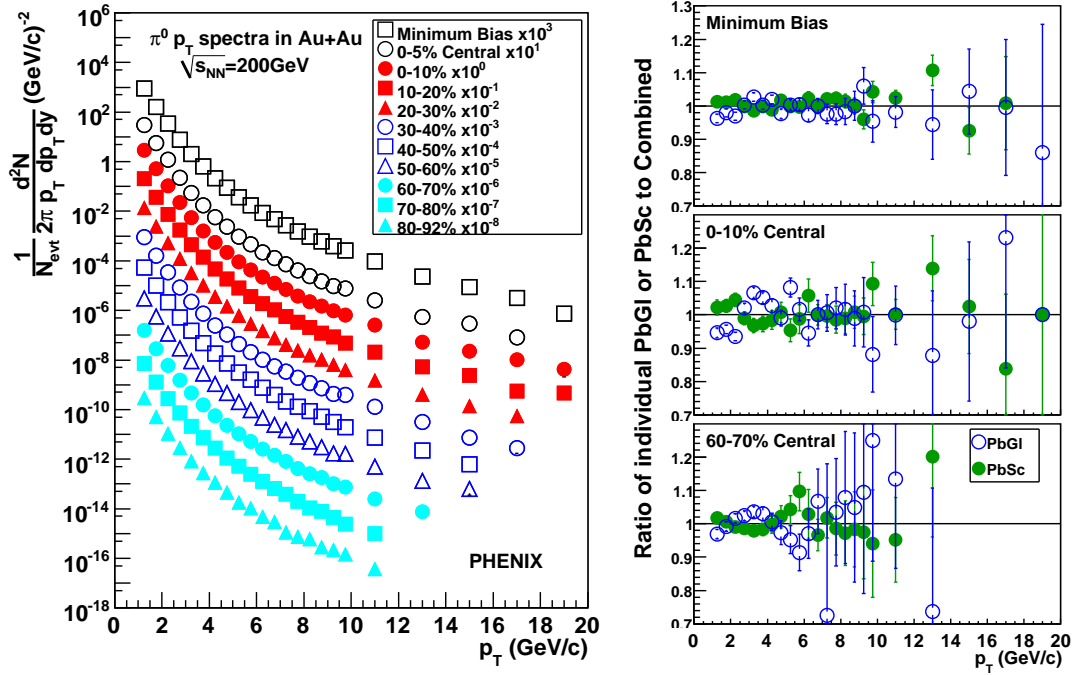}
\caption{\label{fig:spectra} Left: $\pi^0$ invariant yields
for different centralities 
(PbSc and PbGl combined).  Right: consistency between the results obtained
separately from PbSc and PbGl 
}
\end{figure*}

Neutral pions were measured in the
$\pi^{0}\rightarrow\gamma\gamma$ decay channel with the photons
reconstructed in the  Electromagnetic Calorimeter (EMCal) located
in the two central arms of PHENIX ($|\eta|\leq0.35$). The
EMCal~\cite{Aphecetche:2003zr} 
consists of two subsystems: six sectors of lead-scintillator sandwich
calorimeter (PbSc) and two sectors of lead-glass \v{C}erenkov
calorimeter (PbGl) at the radial distance of about 5\,m.
The fine segmentation of the EMCal ($\delta\phi\times\delta\eta$
$\sim0.01\times0.01$ for 
PbSc and $\sim0.008\times0.008$ for PbGl) ensures that the
two photons from a $\pi^{0}\rightarrow\gamma\gamma$ decay are well 
resolved up to $p_{\rm T}^{\pi^0} \approx$ 12 (PbSc) and 16 (PbGl)\,GeV/$c$.
Data from the two subsystems were analyzed separately and the
fully corrected results were combined.

Details of the analysis including extraction of the raw $\pi^0$ yield,
correction for acceptance, detector response (energy resolution, dead
areas), reconstruction efficiency (particle identification cuts) have
been described elsewhere~\cite{Adler:2006bw,ppg079}.
In this analysis the higher $p_{\rm T}$ range required correction for losses
in the observed (raw) $\pi^0$s due to ``cluster merging''. 


%
%

With increasing $\pi^0$ momentum, the minimum opening angle of 
the two decay photons decreases, and eventually
they will be reconstructed as
a single cluster.  Such ``merging'' reaches 50\% of the total raw yield at
$p_{\rm T}$=14~GeV/$c$ in the PbSc and at $p_{\rm T}$=18~GeV/$c$ in the PbGl due
to their different granularity and Moli\`radius.
Merged clusters were rejected by various shower profile cuts, and the loss 
was determined by simulated single $\pi^0$s embedded into real events
and analyzed with the same cuts.
The loss increases slowly with centrality.
The systematic uncertainties
were estimated by comparing $\pi^0$ yields in the PbSc extracted
in different windows of asymmetry 
$|E_{\gamma_1}-E_{\gamma_2}|/(E_{\gamma_1}+E_{\gamma_2})$
and also by comparing yields in the PbSc and PbGl.

%
%

\begin{table}
\caption{\label{tab:syst}
Summary of the systematic uncertainties on the
$\pi^0$ yield extracted independently with the PbSc (PbGl) electromagnetic
calorimeters.  All but one, off-vertex $\pi^0$, are uncorrelated
between PbSc and PbGl, and centrality dependence is negligible. The
last row is the total systematic uncertainty on the combined spectra. Detailed
description of how the errors are correlated as a function of $p_{\rm T}$ can be
found in~\protect\cite{ppg079}. }
\begin{ruledtabular} \begin{tabular}{lcccccccc}
$p_{\rm T}$ (GeV/$c$) &  2 & 6  & 10 & 16 \\  \hline
uncertainty source & \multicolumn{4}{c}{PbSc (PbGl)} \\ \hline
yield extraction (\%) &  3.0 (4.1) & 3.0 (4.1) & 3.0 (4.1) & 3.0 (4.1) \\
PID efficiency (\%) &  3.5 (3.9) & 3.5 (3.5) & 3.5 (3.7) & 3.5 (3.9) \\
Energy scale (\%) & 6.7 (9.0) & 8.0 (9.2) & 8.0 (8.2) & 8.0 (12.3) \\
Acceptance (\%) & 1.5 (4.1) & 1.5 (4.1) & 1.5 (4.1) & 1.5 (4.1) \\
$\pi^0$ merging (\%) & -- (--) & -- (--) & 4.4 (--) & 28 (4.8) \\
Conversion (\%) &  3.0 (2.5) & 3.0 (2.5) & 3.0 (2.5) & 3.0 (2.5) \\
off-vertex $\pi^0$ (\%) &  1.5 (1.5) & 1.5 (1.5) & 1.5 (1.5) & 1.5 (1.5) \\
\hline
Total (\%) &  8.7 (12) &   9.8 (11) &   11 (11) &   30 (15) \\
\hline
PbSc and PbGl & \multirow{2}{0mm}{7.0} & \multirow{2}{0mm}{7.5} & \multirow{2}{0mm}{7.6} & \multirow{2}{0mm}{14} \\ 
combined: Total (\%) & &  &  & \\
\end{tabular} \end{ruledtabular}
\end{table}

We considered two sources of $\pi^0$s not originating from the
collision vertex: those produced in nuclear interactions of hadrons
with detector material (instrumental background) and 
feed-down products from weak decay of higher mass hadrons (physics
background). Based upon simulations both the instrumental background and
feed-down background were found to be negligible
($<$1\% above $p_{\rm T}$$>$$2.0$\,GeV/$c$) except for the contribution from 
$K^0_S$ decay ($\approx$3\% of $\pi^0$ yield for $p_{\rm T}$$>$1\,GeV/$c$),
which has been subtracted from the data. Finally the yields were corrected
to the center of the $p_{\rm T}$ bins using the local slope.

%
%
The main sources of systematic uncertainties are yield extraction, 
efficiency corrections, and energy scale, none of which exhibit
a significant centrality dependence.  The PbSc and PbGl 
detectors have quite different systematics with all but one of them
(off-vertex $\pi^0$) uncorrelated.  Therefore, when combining their
results, the total error is
reduced in the weighted average of the two independent measurements.
The final systematic uncertainties (one standard deviation) on the
spectra are shown in Table~\ref{tab:syst}.

%
%


\begin{figure}[t]
\includegraphics[width=1.0\linewidth]{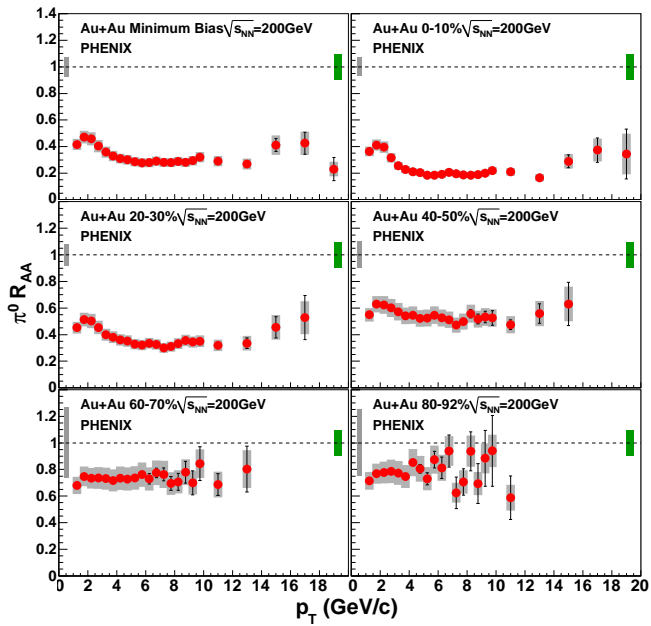}
\caption{\label{fig:RAA} Nuclear modification factor ($R_{\rm AA}$) for
$\pi^0$s.  Error bars
are statistical and $p_{\rm T}$-uncorrelated errors, boxes around the points
indicate $p_{\rm T}$-correlated errors.  Single box around $R_{\rm AA}$=1 on 
the left is the 
error due to $N_{\rm coll}$, whereas the single box on the right is the
overall normalization error of the p+p reference spectrum.
}
\end{figure}

\begin{figure}[htb]
\includegraphics[width=1.0\linewidth]{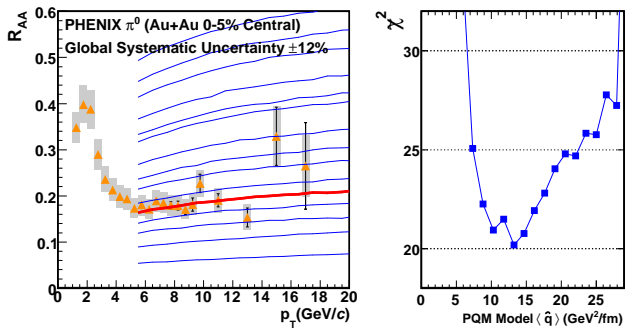}
\caption{\label{fig:fig4} 
Left: $\pi^0$ $R_{\rm AA}$ for the most
    central (0-5\%) Au+Au collisions and PQM model calculations for
    different values of $\mean{\hat{q}}$.  Red curve: best fit.
    Right: $\tilde{\chi}^2(\epsilon_{b}, \epsilon_{c}, {p})$ 
    distribution for a wide range of values of $\mean{\hat{q}}$. 
}
\end{figure}

\begin{figure}[thb]
\includegraphics[width=1.0\linewidth]{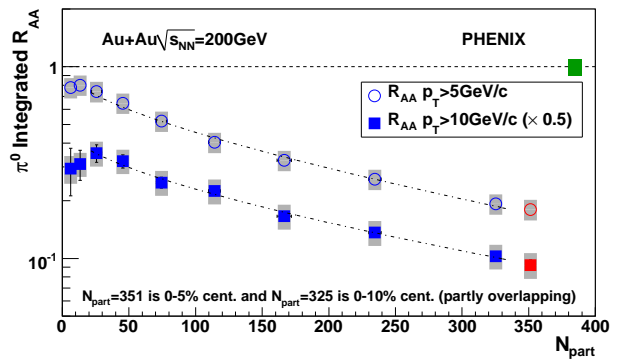}
\caption{\label{fig:int_raa} Integrated nuclear modification factor
($R_{\rm AA}$) for $\pi^0$ 
as a function of collision centrality expressed in terms of
$N_{\rm part}$. The error bars/bands are the same as in Fig.~\ref{fig:RAA}.
The last two points correspond to overlapping centrality bins,
0-10\% and 0-5\%. The dashed lines show the fit to a function. See text.}
\end{figure}

%
%

The left panel of Figure~\ref{fig:spectra} shows the $\pi^0$ invariant yield 
spectra for all centralities as well as minimum bias, combined from
the independent PbSc and PbGl measurements.
In the overlap region the results are consistent with those published
earlier~\cite{Adler:2006bw} while the errors are reduced by a factor of 
2 to 2.5.
The right panel shows the ratios of
PbSc and PbGl spectra to the combined one at three centralities.
The spectra are quite similar at all centralities: 
when fitting $p_{\rm T}$$>$5\,GeV/$c$ with a power law function ($\propto p_{\rm T}^n$), 
the exponents vary from $n=-8.00\pm0.12$ in 0-5\% to $n=-8.06\pm0.08$ in the
80-92\% (most peripheral) bin.  Note that $n=-8.22\pm0.09$ in p+p collisions.
The errors are combined statistical errors and systematic uncertainties.

To quantify the comparison of spectra in heavy ion and p+p collisions,
the nuclear modification factor ($R_{\rm AA}$)
\[R_{\rm AA} = \frac{1/{N_{\rm evt}} dN/dydp_{\rm T}}{\mean{T_{\rm AB}} d\sigma_{pp}/dydp_{\rm T}} \]
is used 
where $\sigma_{pp}$ is the production cross section of the particle in
p+p collisions, and $\mean{T_{\rm AB}}$ is the nuclear thickness function
averaged over a range of impact parameters for the given centrality,
calculated within a Glauber model~\cite{Miller:2007ri}.
 Figure~\ref{fig:RAA} shows $R_{\rm AA}$ for $\pi^0$ at different
   centralities, the 0-5\% bin is shown on Figure~\ref{fig:fig4}.
The reference p+p yield was obtained from the 2005 (Run-5) RHIC p+p
measurement~\cite{Adare:2007dg}.

$R_{\rm AA}$ reaches $\sim$0.2 in 0-10\,\% centrality at
$p_{\rm T}$$>$5\,GeV/$c$ with very little (if any) $p_{\rm T}$ dependence.
This trend is compatible with most current energy loss
models but not with a semi-opaque medium assumption,
where $R_{\rm AA}$ would decrease with increasing $p_{\rm T}$ \cite{Renk:2006pr}.
While its magnitude changes, the suppression pattern itself
is remarkably similar at all centralities suggesting that the bulk
$R_{\rm AA}$ (integrated over the azimuthal angle) 
is sensitive only to the $N_{\rm part}$
but not to the specific geometry.  
Consequently, study of the 
$p_{\rm T}$-integrated $R_{\rm AA}$ {\it vs.} centrality is instructive.

Figure~\ref{fig:int_raa} shows the integrated nuclear modification
factor ($p_{\rm T}$$>$5\,GeV/$c$, and $p_{\rm T}$$>$10\,GeV/$c$) for $\pi^0$s as a
function of centrality, with the last two points indicating
overlapping 0-10\% and 0-5\% bins. 
In both cases the suppression increases monotonically
with $N_{\rm part}$ without any sign of saturation, suggesting that larger
colliding systems (such as U+U planned at RHIC) should exhibit even 
more suppression.

The common power-law behavior ($\propto p_{\rm T}^n$) in p+p and Au+Au
allows the suppression to be re-interpreted as a fractional energy loss
$S_{\rm loss}=1-R_{\rm AA}^{1/(n-2)}$ where $n$ is the power-law exponent,
and we found that $S_{\rm loss}\propto N_{\rm part}^a$~\cite{Adler:2006bw}.
Fitting the integrated $R_{\rm AA}$ with a function
$R_{\rm AA}=(1-S_{0} N_{\rm part}^{a})^{n-2}$ gives $a=0.58\pm0.07$ for
$N_{\rm part}$$>$20 for $p_{\rm T}$$>$5\,GeV/$c$, and $a=0.56\pm0.10$ for
$p_{\rm T}$$>$10\,GeV/$c$. The GLV~\cite{Gyulassy:2000fs} and
PQM~\cite{Loizides:2006cs} models predict that $a\approx 2/3$,
which is consistent with the data.
The fitted values of $S_0$ are (8.3$\pm$3.3)$\times10^{-3}$
and (9.2$\pm$4.9)$\times10^{-3}$ for
$p_{\rm T}$$>$5\,GeV/$c$ and $p_{\rm T}$$>$10\,GeV/$c$, respectively.
The fits are shown as dashed lines in Fig.~\ref{fig:int_raa}.
Note that in this interpretation a constant $S_{\rm loss}$ 
(independent of $p_{\rm T}$) implies that the energy loss increases
with $p_{\rm T}$.

   We use the highest centrality (0--5\%) $R_{\rm AA}$ data as shown on 
Fig.~\ref{fig:fig4} to constrain the PQM model parameters. The procedure 
is described in detail in~\cite{ppg079}. First we break up the errors 
of the measured points into Type A ($p_{\rm T}$-uncorrelated, 
statistical $\oplus$ systematic, $\sigma_i$), Type B 
($p_{\rm T}$-correlated, $\sigma_{b_i}$, boxes on Fig.~2) and
Type C (normalization, uniform fractional shift for all points, 
$\sigma_c$). Then taking the theory curves calculated for different 
values of the input parameter $p$, one would normally perform a 
least-squares fit to the theory by finding the values of 
${p}$, ${\epsilon}_b$, ${\epsilon}_c$ that minimize:   
\begin{equation}
\chi^2={\left[\sum_{i=1}^{n}
{{(y_i+\epsilon_b \sigma_{b_i} +\epsilon_c y_i \sigma_c -\mu_i({p}))^2}  \over {\sigma_i^2}}+ {\epsilon_b^2 }+{\epsilon_c^2 }\right]} \qquad ,
\label{eq:-2lnL-gaussian-sys}
\end{equation}
where $\epsilon_b$ and $\epsilon_c$ are the fractions of the type B 
and C systematic uncertainties that all points are displaced together. 
It is important to note that Eq.~\ref{eq:-2lnL-gaussian-sys} follows 
the $\chi^2$-distribution with $n+2$ degrees of freedom when
$p$, $\epsilon_b$ and $\epsilon_c$ are fixed, because it is the sum
of $n+2$ independent Gaussian distributed random variables. 

However, for the present data, the statistical and random systematic 
uncertainties are such that the shift in the measurement $y_i$ due to 
the correlated systematic uncertainties preserves the fractional 
type A uncertainty. Thus, we use a least squares fit of the quantity 
$\tilde{\chi}^2$ to estimate the best fit parameters, where 
$\tilde{\chi}^2$ is Eq.~\ref{eq:-2lnL-gaussian-sys} with $\sigma_i$ 
replaced by 
$\tilde{\sigma}_i=\sigma_i ({y_i+\epsilon_b \sigma_{b_i} 
+\epsilon_c y_i \sigma_c})/{y_i}$, 
which is the uncertainty scaled by the multiplicative shift in $y_i$ 
such that the fractional uncertainty is unchanged under shifts. For 
any fixed values of $\epsilon_b$, $\epsilon_c$, $\tilde{\chi}^2$ still 
follows the $\chi^2$ distribution with $n+2$ degrees of freedom. 
The best fit, the minimum of $\tilde{\chi}^2$,  is found by standard 
methods (for example using a MINUIT type minimization algorithm) and 
the correlated uncertainties of the best fit parameters are estimated 
in the Gaussian approximation by 
$\tilde{\chi}^2(\epsilon_{b}, \epsilon_{c}, 
{p})=\tilde{\chi}^2_{\rm min}+N^2$ for $N$ 
standard deviation uncertainties. The right panel of Fig.~4 shows the 
$\tilde{\chi}^2(\epsilon_{b}, \epsilon_{c}, {p})$ 
distribution for a wide range of values of the PQM model parameter
$\mean{\hat{q}}$. Our data constrain the PQM model transport
coefficient  $\mean{\hat{q}}$ as
13.2 $^{+2.1}_{-3.2}$ and $^{+6.3}_{-5.2}$\,GeV$^2$/fm at the one and
two standard deviation levels.  
These constraints include only the
experimental uncertainties and do not account for the large model dependent
differences in the quenching scenario and description of the medium. 
Extracting
fundamental model-independent properties of the medium from the present data
requires resolution of ambiguities and open questions in the models 
themselves which also will
have to account simultaneously for the $p_{\rm T}$ and centrality (average
pathlength) dependence.
This work demonstrates the
power of data for pion production in constraining the energy loss of
partons.
The data can be fitted with a constant in the entire
$p_{\rm T}$$>5$\,GeV/$c$ range as well: the slope of a simple linear fit
is 0.0017 $^{+0.0035}_{-0.0039}$ and $^{+0.0070}_{-0.0076}$\,$c$/GeV 
at the one and two standard deviation levels.

%
%

%
%
In summary, PHENIX has measured neutral pions in Au+Au collisions at
$\sqrt{s_{NN}}$=200\,GeV at mid rapidity in the transverse momentum
range of 1$<$$p_{\rm T}$$<$20\,GeV/$c$, analyzing high statistics RHIC Run-4 run
data. 
The shape of the spectra is similar for all centralities, as is the
shape of $R_{\rm AA}(p_{\rm T})$ at $p_{\rm T}$$>$5GeV/$c$.
In central collisions the yield is suppressed by a factor of $\sim$5 at
5\,GeV/$c$ compared to the binary scaled p+p reference 
and the suppression prevails with little or no change up to 20\,GeV/$c$.
Studying the integrated $R_{\rm AA}$ {\it vs.} centrality 
we find that it does not saturate at
this nuclear size; also the prediction 
$S_{\rm loss} \propto N_{\rm part}^{2/3}$\cite{Gyulassy:2000fs,Loizides:2006cs}
is consistent with our data
and in this picture the energy loss increases with $p_{\rm T}$.  
Using the 0-5\% (most central) $R_{\rm AA}$ we find that
the transport coefficient $\mean{\hat{q}}$ of the PQM model is constrained to
13.2 $^{+2.1}_{-3.2}$ ($^{+6.3}_{-5.2}$)\,GeV$^2$/fm at the one (two)
$\sigma$ level.  A simple linear fit with zero slope
is also consistent with our data.



We thank the staff of the Collider-Accelerator and 
Physics Departments at BNL for their vital contributions.  
We acknowledge support from 
the Office of Nuclear Physics in DOE Office of Science and NSF (U.S.A.), 
MEXT and JSPS (Japan), 
CNPq and FAPESP (Brazil), 
NSFC (China), 
MSMT (Czech Republic),
IN2P3/CNRS, and CEA (France), 
BMBF, DAAD, and AvH (Germany), 
OTKA (Hungary), 
DAE (India), 
ISF (Israel), 
KRF and KOSEF (Korea), 
MES, RAS, and FAAE (Russia),
VR and KAW (Sweden), 
U.S. CRDF for the FSU, 
US-Hungarian NSF-OTKA-MTA, 
and US-Israel BSF.



\end{document}